\documentclass[%
nofootinbib,
notitlepage,
amsfonts,
amsmath,
amssymb,
twocolumn,
pra,
superscriptadress,
longbibliography
]{revtex4-2}
\usepackage{dcolumn}

\usepackage{mathptmx}

\usepackage[T1]{fontenc}
\usepackage{graphicx}
\usepackage{dcolumn}
\usepackage{bm}
\usepackage{braket}
\usepackage{mathptmx}
\usepackage[normalem]{ulem}
\usepackage{color}
\usepackage{xcolor}

\begin{document}

\title{High-threshold fault-tolerant quantum computation with the Gottesman--Kitaev--Preskill qubit\\ 
under noise in an optical setup
}
\author{Kosuke Fukui} 
\affiliation{%
Department of Applied Physics, School of Engineering, The University of Tokyo,
7-3-1 Hongo, Bunkyo-ku, Tokyo 113-8656, Japan}
\renewcommand{\baselinestretch}{1.0} 
\setlength\textfloatsep{20pt}
\setlength\abovecaptionskip{0pt}

\begin{abstract}
To implement fault-tolerant quantum computation (FTQC) with continuous variables, continuous variables need to be digitized using an appropriate code such as the Gottesman--Kitaev--Preskill (GKP) qubit. 
The scheme introduced in [K. Fukui {\it et. al.}, Phys. Rev. X {\bf 8}, 021054 (2018)] has reduced the threshold of a squeezing level required for continuous-variable FTQC to less than 10 dB, assuming noise derived from the GKP qubit itself.
In this work, we propose a scheme to improve noise tolerance during the construction of large-scale cluster state used for FTQC with the GKP qubits.
In our scheme, a small-scale cluster state is prepared by employing a maximum-likelihood method, the entanglement generation via the Bell measurement, and a probabilistic reliable measurement.
Then, a large-scale cluster state is construct from the small-scale cluster states via the encoded Bell measurement.
In the numerical calculations, we assume errors derived from the two-mode gate and loss in the homodyne measurement in addition to noise from the GKP qubit itself.
The results show that the thresholds of a squeezing level are around 8.1, 9.6, and 12.4 dB for loss in the homodyne measurement 0, 5, and 10~\%, respectively.
Hence, this work provides a way toward continuous-variable FTQC with a feasible squeezing level.
\end{abstract}
                           
\maketitle

\section{Introduction}\label{Intro}
Quantum computation (QC) has a great deal of potential to efficiently solve some hard problems for conventional computers~\cite{shor1999polynomial, grover1997quantum}. 
To realize large-scale QC, a continuous-variable system is a promising platform~\cite{takeda2019toward,pfister2019continuous,fukui2022building}; in fact, lager-scale cluster states composed of the squeezed vacuum states has been experimentally generated in an optical setup~\cite{yokoyama2013ultra,yoshikawa2016invited,asavanant2019generation, larsen2019deterministic,larsen2021deterministic,asavanant2021time}. 
Furthermore, more than thousands of a frequency-encoded cluster state in an optical set up has been generated \cite{pysher2011parallel,chen2014experimental,roslund2014wavelength,cai2017multimode}.
In addition to an optical setup, the CV system a circuit QED~\cite{grimsmo2017squeezing}, opto-mechanics~\cite{schmidt2012optomechanical,houhou2015generation}, atomic ensembles~\cite{ikeda2013deterministic,motes2017encoding}, and a trapped ion mechanical oscillator~\cite{fluhmann2018sequential, fluhmann2019encoding} are also promising candidates for large-scale QC with CVs.

Towards continuous-variable fault-tolerant QC (CV-FTQC), it is known that CVs need to be encoded into appropriate bosonic codes~\cite{pegg1989phase,chuang1997bosonic,albert2018performance,grimsmo2020quantum}, such as a cat code~\cite{cochrane1999macroscopically}, a binomial code~\cite{michael2016new}, or the Gottesman--Kitaev--Preskill (GKP) code~\cite{gottesman2001encoding} referred to as the GKP qubit in this work. 
This is because the squeezed vacuum state can not handle the accumulation of analog errors, for example, e.g., the Gaussian quantum channel~\cite{gottesman2001encoding} and a photon loss during QC.
In 2014, Menicucci showed the threshold of the squeezing level for CV-FTQC~\cite{menicucci2014fault}, where the GKP qubit is used to perform the quantum error correction during for measurement-based QC (MBQC). 
Recently, there have been many efforts towards CV-FTQC with the GKP qubit~\cite{fukui2017analog,fukui2018high,douce2019probabilistic,vuillot2019quantum,baragiola2019all,shi2019fault,walshe2019robust,pantaleoni2020modular,walshe2020continuous,pantaleoni2021subsystem,grimsmo2021quantum,fukui2018tracking,vuillot2019quantum,
noh2020fault,noh2020encoding,yamasaki2020polylog, noh2021low,tzitrin2021fault,seshadreesan2022coherent,stafford2022biased}, and a promising architecture for a scalable quantum circuit incorporating the GKP qubit~\cite{takeda2017universal,alexander2018universal,bourassa2021blueprint}.
Furthermore, the GKP qubit is a promising element to a variety of quantum information processing such as long-distance quantum communication~\cite{fukui2021all,rozpkedek2021quantum}.

In Ref.~\cite{fukui2018high}, the required squeezing level for CV-FTQC has been alleviated to less than 10 dB, which is within the reach of the current experimental technology~\cite{terhal2016encoding}. 
In an ion trap~\cite{fluhmann2019encoding} and superconducting circuit quantum electrodynamics~\cite{campagne2020quantum}
the GKP qubit has been generated recently in an ion trap~\cite{fluhmann2019encoding} and superconducting circuit quantum electrodynamics~\cite{campagne2020quantum} with an achievable squeezing level close to 10 dB.
In an optical setup, while there are many efforts to develop various ways to generate the GKP qubit~\cite{pirandola2004constructing, pirandola2006continuous, pirandola2006generating, motes2017encoding,eaton2019non,su2019conversion,
arrazola2019machine,tzitrin2020progress,lin2020encoding,hastrup2021generation,fukui2022generating,fukui2021efficient,takase2022gaussian}, the optical GKP qubit has not been generated yet due to the difficulty to obtain a nonlinearity.
Thus, there is a demand to reduce the experimental requirements to generate sufficient GKP qubit to FTQC.
In addition, a noise model in Ref.~\cite{fukui2018high} has assumed that the deviation is derived from the GKP qubit itself.
Considering a practical optical setup, there are additional noises such as imperfection derived from the two-mode gate and the homodyne measurement.
This leads to the degradation of the threshold of a squeezing level, which increases the requirement of CV-FTQC.
For the above reasons, a further reduction of the threshold is needed for CV-FTQC in an optical setup.

In this work, we propose a scheme to improve the required squeezing level for CV-FTQC under noise in the two-mode gate and the homodyne measurement, developing the method to implement the highly-reliable construction of the large-scale cluster state by harnessing analog information contained in the GKP qubits.
Specifically, our method consists of two parts.
One is to make use of the Gauss-Markov theorem, which is widely known in statistics, to reduce the noise of the GKP qubits in the construction of the small-scale cluster state.
The other is the reliable deterministic entanglement generation to construct the large-scale cluster state from the small-scale cluster states. 
In this operation, we select the most reliable entanglement between node qubits by using a maximum-likelihood method, which allows us to safely remove the entanglements except for the most reliable one by using the repetition code.
Accordingly, the required squeezing level for CV-FTQC using the proposed method can be reduced to 8.1, 9.6, and 12.4 dB for the transmission loss in the homodyne measurement $l =0$, 5, and 10~$\%$, respectively, assuming the imperfection of the two-mode gate.

The rest of the paper is organized as follows. 
In Sec.~\ref{Sec2}, we briefly review the background knowledge regarding the GKP qubit, and noise in the two-mode gate and the homodyne measurement. 
In Sec.~\ref{Sec3}, we propose the method to reduce the required squeezing level for CV-FTQC with noise considered in this work.  
In Sec.~\ref{Sec4}, the numerical results show the improvement of the threshold of a squeezing level compared with the conventional methods. Section~\ref{Sec5} is devoted to discussion and conclusion.

\section{Background}\label{Sec2}
In this section, we firstly review the GKP qubits and noise model in this work, assuming three noise sources, e.g., the deviation from the GKP qubit itself, the imperfection of the two-mode gate, and loss in the homodyne measurement.
Then, we describe two techniques used to improve noise tolerance with the GKP qubit: (1) the highly-reliable measurement and (2) the single-qubit level QEC with a maximum-likelihood estimation.
\subsection{The GKP qubit}\label{Sec2A}
Gottesman, Kitaev, and Preskill proposed a method to encode a qubit in an oscillator's $q$ (position) and $p$ (momentum) quadratures to correct errors caused by a small deviation in the $q$ and $p$ quadratures~\cite{gottesman2001encoding}. 
We refer the state encoded by their scheme as the GKP qubit.
The ideal code states of the GKP qubit are Dirac combs in $q$ and $p$ quadratures, which are described as
$\ket {{0}}_{\rm GKP}= \sum_{m=- \infty}^{\infty}\ket{2m\sqrt{\pi}}_q$ and $\ket {{1}}_{\rm GKP}= \sum_{m=- \infty}^{\infty}\ket{(2m+1)\sqrt{\pi}}_q$, 
respectively.
The ideal GKP code state is not a normalizable state and it requires infinite squeezing.
Thus, physical states for the GKP code are finitely squeezed approximations.
The basis of the GKP qubit with finite squeezing is composed of a series of Gaussian peaks of width $\sigma$ and separation $\sqrt{\pi}$ embedded in a larger Gaussian envelope of width 1/$\sigma$. 
The approximate code states $\ket {\widetilde{0}}$ and $\ket {\widetilde{1}}$ are defined as  
\begin{eqnarray}
\ket {\widetilde{0}} &\propto &   \sum_{t=- \infty}^{\infty} \int \mathrm{e}^{-2\pi\sigma^2t^2}\mathrm{e}^{-(q-2t\sqrt{\pi})^2/(2\sigma^2)}\ket{q}  dq,  \\
\ket {\widetilde{1}} &\propto & \sum_{t=- \infty}^{\infty} \int \mathrm{e}^{-\pi\sigma^2(2t+1)^2/2}  
\mathrm{e}^{-(q-(2t+1)\sqrt{\pi})^2/(2\sigma^2)}\ket{q}  dq .     
\end{eqnarray}
The squeezing level $s$ is defined by $s=-10{\rm log}_{10}2{\sigma}^2$.
In the case of finite squeezing, there is a finite probability of misidentifying $\ket {\widetilde{0}}$ as $\ket {\widetilde{1}}$, and vice versa. 
Provided the magnitude of the true deviation is more than $\sqrt{\pi}/2$ from the peak value, the decision of the bit value is incorrect. 
The probability $E(\sigma^2)$ of misidentifying the bit value is calculated by
\begin{equation}
E(\sigma^2) = 1-\int_{\frac{-\sqrt{\pi}}{2}}^{\frac{\sqrt{\pi}}{2}} dx \frac{1}{\sqrt{2\pi {\sigma} ^2}} e^{-\frac{x^2}{2{\sigma} ^2}},
\label{eq3}
\end{equation}
which corresponds to bit- or phase-flip errors on the GKP qubit.
We mention that $q$ and $p$ quadratures are also referred as $Z$ and $X$ bases, respectively.

We also describe the so-called qunaught state which is introduced in Ref.~\cite{duivenvoorden2017single} for quantum sensing applications. The qunaught state $\ket{\varnothing}$ is described as
\begin{align}
     \ket{\varnothing} \propto \sum_{k=-\infty}^{\infty} e^{-i\sqrt{2\pi}k\hat{p}} \ket{0}_{q}
     = \sum_{k=-\infty}^{\infty} e^{i \sqrt{2\pi}k\hat{q}} \ket{0}_{p}.
 \end{align}
Two qunaught states are transformed to a Bell pair of the GKP qubits by a 50:50 beam-splitter coupling as
$ \ket{\varnothing}\otimes  \ket{\varnothing} \mapsto ({\ket{\bar{0}\bar{0}}+\ket{\bar{1}\bar{1}}})/{\sqrt{2}}$~\cite{walshe2020continuous}.
In the case of a qunaught state with the variances $\sigma^2$ in both quadratures, the variances of each GKP qubit of the generated Bell pair are $\sigma^2$ in both quadratures.
One of the advantages to employ the Bell pair from two qunaught states is error tolerance in terms of variances. 
Specifically, the variances of each GKP qubit in the entangled pair generated from CZ gate between two GKP qubits are $\sigma^2$ and $2\sigma^2$ in the $q$ and $p$ quadratures, respectively.
Thus, the error probability for the entangled pair from two qunaught states are smaller than that prepared from the two GKP qubits, which are obtained from Eq.~\ref{eq3}.
In this work, we prepare the small-scale cluster state using qunaught states, as described in Sec.~\ref{Sec3A}.

\subsection{Noise in the two-mode gate}\label{Sec2B}
For the noisy two-mode gate with CVs, we consider the Controlled-NOT (CX) gate demonstrated in Refs.~\cite{yoshikawa2008demonstration, shiozawa2018quantum}, which is also called the optical quantum Nondemolition (QND) gate.
In the QND gate, the interaction between the control qubit C and the target qubit T is realized by using a beam-splitter coupling with two ancillary squeezed vacuum states~\cite{yoshikawa2008demonstration, shiozawa2018quantum}, assuming that $\sqrt{R}$ is a transmittance coefficient for a beam-splitter coupling is $R$ and the squeezing parameter for the ancillary state A(B) is $r_{\rm A}~(r_{\rm B})$.
The QND gate transforms the quadrature operators for the control and target qubits as
\begin{eqnarray}
 \hat{q}_{\rm C} &\to &   \hat{q}_{\rm C}- \sqrt{\frac{1-R}{1+R}} \hat{q}_{\rm A}  \mathrm{e}^{-r_{\rm A}}, \\
 \hat{p}_{\rm C} &\to & \hat{p}_{\rm C} -\frac{1-R}{\sqrt{R}} \hat{p}_{\rm T}+ \sqrt{\frac{R(1-R)}{1+R}}\hat{p}_{\rm B} \mathrm{e}^{-r_{\rm B}} ,  \label{eq5} \\
 \hat{q}_{\rm T} &\to &  \frac{1-R}{\sqrt{R}}\hat{q}_{\rm C}+ \hat{q}_{\rm T}+ \sqrt{\frac{R(1-R)}{1+R}}\hat{q}_{\rm A} \mathrm{e}^{-r_{\rm A}} , \label{eq6}\\
\hat{p}_{\rm T} &\to &  \hat{p}_{\rm T}+ \sqrt{\frac{1-R}{1+R}}\hat{p}_{\rm B} \mathrm{e}^{-r_{\rm B}},
\end{eqnarray}
where $\hat{q}_{\rm C} (\hat{p}_{\rm C})$, $\hat{q}_{\rm T} (\hat{p}_{\rm T})$, $\hat{q}_{\rm A} (\hat{p}_{\rm A})$, and $\hat{q}_{\rm B} (\hat{p}_{\rm B})$ are the quadrature operators of the control qubit, target qubit, and two squeezed vacuum states in the $q$ ($p$), respectively. 
In the case that the coefficient $({1-R})/{\sqrt{R}}$ is equal to 1 and the squeezing level of ancillary squeezed vacuum states is infinite, the QND gate is equivalent to the ideal CX gate, which corresponds to the operator exp(-$i\hat{q}_{\rm C}\hat{p}_{\rm T}$). For the CZ gate, the QND gate is equivalent to the CZ gate up to local Fourier transformations.

Regarding the variance of the GKP qubit, the QND gate changes the variances of the control and target qubits as
\begin{eqnarray}
\sigma^2_{{\rm C},q} &\to& \sigma^2_{{\rm C},q}+\frac{1-R}{1+R}\sigma^2_{A}, \\
\sigma^2_{{\rm C},p} &\to& \sigma^2_{{\rm C},p}+\frac{(1-R)^2}{R}\sigma^2_{{\rm C},p}+\frac{R(1-R)}{1+R}\sigma^2_{B},\\
\sigma^2_{{\rm T},q} &\to& \sigma^2_{{\rm C},q}+\frac{(1-R)^2}{R}\sigma^2_{{\rm T},q}+\frac{R(1-R)}{1+R}\sigma^2_{A}, \\
\sigma^2_{{\rm T},p} &\to& \sigma^2_{{\rm T},p}+\frac{1-R}{1+R}\sigma^2_{B},
\end{eqnarray}
where variances $\sigma^2_{{\rm C},q(p)}$ and $\sigma^2_{{\rm T},q(p)}$ are the variances of the control qubit and the target qubit in the $q$($p$) quadrature, respectively, and $\sigma^2_{{\rm A(B)}}=\frac{1}{2}\mathrm{e}^{-2r_{\rm A(B)}}$ is the variance for the ancillary state A~(B).
In this work, we assume that $\sigma^2_{{\rm A}}=\sigma^2_{{\rm B}}$, and a squeezing level of the ancillary squeezed vacuum is set to 15 dB, which is an achievable squeezing level in the experiment of Ref.~\cite{vahlbruch2016detection}.

\subsection{Photon loss in the homodyne masurement}\label{Sec2C}
Secondly, we consider photon loss in the homodyne measurement.
Noise derived from photon loss is described by a beam-splitter coupling between the data qubit and a vacuum state.
The beam-splitter coupling transforms the quadrature operators in the $q$ and $p$ quadratures as 
\begin{eqnarray}
\hat{q} \to \sqrt{\eta}  \hat{q}+\sqrt{1-\eta} \hat{q}_{\rm vac}, \\
\hat{p} \to \sqrt{\eta}  \hat{p}+\sqrt{1-\eta} \hat{p}_{\rm vac}, 
\end{eqnarray}
where $\sqrt{\eta}$ is a transmittance coefficient for a beam-splitter coupling, $\hat{q}_{\rm vac}$ ($\hat{p}_{\rm vac}$) is operators for the vacuum state in the $q(p)$ quadrature. 
The variance of the input state in the $q(p)$ quadrature, ${\sigma^{2}} _{{\rm in},q (p)}$, changes as,
\begin{equation}
{\sigma^{2}} _{{\rm in},q(p) } \to \sigma^2_{{\rm out},q(p)} = {\eta} \sigma^{2}_{{\rm in},q(p)} +\frac{1-{\eta}}{2}.
\end{equation}
In the measurement after loss, the outcome from the homodyne measurement is multiplied by $1/\sqrt{\eta}$ in classical post-processing since the peaks of the GKP qubit is fixed at the integer multiples of $\sqrt{\pi}$ due to the GKP codewords.
Consequently, the probability to misidentify the bit value in the $q$($p$) quadrature is calculated by $E({\sigma^2}'_{{\rm out},q(p)})$ using Eq.~(\ref{eq3}), where ${\sigma^2}'_{{\rm out},q(p)}$ is given by
\begin{equation}
{\sigma^2}'_{{\rm out},q(p)}=\frac{\sigma^2_{{\rm out},q(p)}}{\eta} =\sigma^{2}_{{\rm in},q(p)} +\frac{1-{\eta}}{2\eta}.
\end{equation}

\subsection{ Highly-reliable measurement}\label{Sec2D}
\begin{figure}[t]
\centering \includegraphics[angle=0, width=1.0\columnwidth]{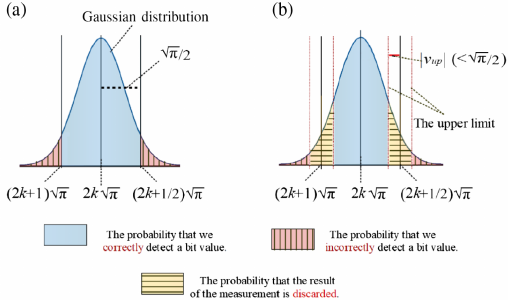} 
     \setlength\abovecaptionskip{0pt}
\caption{Introduction of the highly-reliable measurement. (a) The conventional measurement of the GKP qubit, where the Gaussian distribution followed by the deviation of the GKP qubit whose variance is $\sigma^{2}$. The  plain (blue) region and the region with vertical (red) line represent the different code word $(k-1)$ mod 2 and $(k+1)$ mod 2, respectively. The vertical line regions correspond to the probability of incorrect decision of the bit value. (b) The highly-reliable measurement. The shown dot line represents a upper limit $v_{\rm up}$. The horizontal line areas show the probability that the results of the measurement is discarded by introducing $v_{\rm up}$. The vertical line areas show the probability that our method fails.}
\label{fig:4}
\end{figure}

The highly-reliable measurement (HRM) reduces the probability of misidentifying the bit value of the GKP qubit by introducing upper limit $v_{\rm up}$ as a decision line of the bit value, as shown in Fig.~\ref{fig:4} (a).
In the conventional measurement, the decision sets an upper limit for $|\Delta_{m}|$ at $\sqrt{\pi}/2$, and assigns the bit value $k$ = $(2 t + k)\sqrt{\pi} $. 
In the HRM, the decision sets an upper limit at $v_{\rm up}(<\sqrt{\pi}/2)$ to give the maximum deviation that will not cause incorrect measurement of the bit value as shown in Fig. \ref{fig:4}. If the above condition $|\Delta_{\rm m}| < v_{\rm up}$ is not satisfied, we discard the result. Since the measurement error occurs when $|\bar{\Delta}|$ exceeds $|\sqrt{\pi}/2+v_{\rm up}|$, the error probability decreases as increasing $v_{\rm up}$ at the cost of the success probability of the measurement. 
The probability of misidentifying the bit value with the HRM for the variance $\sigma^2$ is given by
\begin{equation}
E_{v_{\rm up}}(\sigma^2) =\frac{P_{v_{\rm up}}^{\rm in}(\sigma^2) }{{P_{v_{\rm up}}^{\rm cor}(\sigma^2) }+{P_{v_{\rm up}}^{\rm in}}(\sigma^2) }, \label{eq.post}
\end{equation}
where ${P_{v_{\rm up}}^{\rm cor}(\sigma^2)}$ is the probability that the true deviation $|\bar{\Delta}|$ falls in the correct area, and ${P_{v_{\rm up}}^{\rm in}(\sigma^2)}$ is the probability that the true deviation $|\bar{\Delta}|$ falls in the incorrect area. 
The probabilities, ${P_{v_{\rm up}}^{\rm cor}}(\sigma^2)$ and ${P_{v_{\rm up}}^{\rm in}}(\sigma^2)$, are given by
\begin{equation}
P_{v_{\rm up}}^{\rm cor}= \sum_{k=-\infty}^{+\infty} \int_{2k\sqrt{\pi}-\frac{\sqrt{\pi}}{2}+v_{\rm up}}^{2k\sqrt{\pi}+\frac{\sqrt{\pi}}{2}-v_{\rm up}} dx \frac{1}{\sqrt{2\pi {\sigma}^2}}\mathrm{e}^{-\frac{x^2}{{2{\sigma}^2}}}
\end{equation}
and 
\begin{equation}
P_{v_{\rm up}}^{\rm in}= \sum_{k=-\infty}^{+\infty} \int_{(2k+1)\sqrt{\pi}-\frac{\sqrt{\pi}}{2}+v_{\rm up}}^{(2k+1)\sqrt{\pi}+\frac{\sqrt{\pi}}{2}-v_{\rm up}} dx \frac{1}{\sqrt{2\pi {\sigma}^2}}\mathrm{e}^{-\frac{x^2}{{2{\sigma}^2}}},
\end{equation}
respectively.

\subsection{The SQEC with a maximum-likelihood estimation}\label{Sec2E}
In this subsection, we describe the single-qubit level QEC (SQEC) and propose the SQEC with a maximum-likelihood estimation (ME-SQEC).
The SQEC is used to correct a small displacement (deviation) error~\cite{gottesman2001encoding}.
We describe the SQEC in the $q$ quadrature (see also Appendix~\ref{AppA} for the SQEC in the $p$ quadrature).
To correct the small deviation of the data qubit D in the $q$ quadrature, the ancilla qubit A is prepared in the logical state $\ket {\widetilde{+}}_{\rm A}$, and is entangled with the data qubit by the ideal CX gate, where the data and the ancilla qubits are the target and control qubits, respectively.
The ideal CX gate described by exp(-$i\hat{q}_{\rm a}\hat{p}_{\rm D}$) transforms the deviation values of data and ancilla qubits as
\begin{eqnarray}
\overline{\Delta}_{\rm {\it q},a} &\to & \overline{\Delta}_{\rm {\it q},a}+ \overline{\Delta}_{\rm {\it q},D},\\ 
\overline{\Delta}_{\rm {\it p},a} &\to& \overline{\Delta}_{\rm {\it p},a}, \\
\overline{\Delta}_{\rm {\it q},D} &\to& \overline{\Delta}_{\rm {\it q},D},\\  
\overline{\Delta}_{\rm {\it p},D} &\to& \overline{\Delta}_{\rm {\it p},D}-\overline{\Delta}_{\rm {\it p},a},
\end{eqnarray}
where $\overline{\Delta}_{\rm {\it q},D} ( \overline{\Delta}_{\rm {\it p},D})$ and $\overline{\Delta}_{\rm {\it q},a}  (\overline{\Delta}_{\rm {\it p},a} )$ are the true deviation values of the data and ancilla qubits in the $q$ ($p$) quadrature, respectively.
Then we measure the ancilla qubit in the $q$ quadrature, and obtain the measurement outcome $m_{q,{\rm m}}=(2 t + k)\sqrt{\pi}+{\Delta}_{\rm m{\it q}, a}$ to minimize $|{\Delta}_{\rm m{\it q}, a}|$, where $k$ is the bit value and $t = 0, \pm 1, \pm 2,\cdots$.
Then, we perform the displacement operation on the data qubit in the $q$ quadrature by the measured deviation${\Delta}_{\rm m{\it q}, a}$. 
If $| \overline{\Delta}_{\rm {\it q},a}+ \overline{\Delta}_{\rm {\it q},D}|$ is less than $\sqrt{\pi}/2$, the true deviation value of the data qubit in the $q$ quadrature changes to $-\overline{\Delta}_{\rm {\it q},a}$, where 
the deviation of the data qubit, $\overline{\Delta}_{\rm {\it q},D}$, is displaced by the measured deviation $\overline{\Delta}_{\rm {\it q},a}+ \overline{\Delta}_{\rm {\it q},D}$. 
On the other hand, if $ |\overline{\Delta}_{\rm {\it q},a}+ \overline{\Delta}_{\rm {\it q},D}|$ is more than $\sqrt{\pi}/2$, the bit-flip error occurs.
Using Eq.~(\ref{eq3}), the error probability of the bit-flip error is obtained by $E(\sigma^2_{{\rm D},q}+\sigma^2_{{\rm a},q})$, assuming that the variances of the data (ancilla) qubit in the $q$ and $p$ quadratures are $\sigma^2_{{\rm D(a)},q}$ and $\sigma^2_{{\rm D(a)},p}$, respectively.
As a consequence, the SQEC in the $q$ quadrature reduces the variance of the data qubit in the $q$ quadrature from $\sigma^2_{{\rm D},q}$ to ${\sigma^2_{{\rm a},q}}$, when $\sigma^2_{{\rm D},q}>{\sigma^2_{{\rm a},q}}$.
Regarding the variance of the data qubit in the $p$ quadrature, the SQEC increases the variance from $\sigma^2_{{\rm D},p}$ to $\sigma^2_{{\rm D},p}+{\sigma^2_{{\rm a},p}}$

In this work, we introduce the SQEC with a maximum-likelihood estimation (ME-SQEC) to improve noise tolerance.
In the ME-SQEC, we estimate the true deviation of the data qubit by considering the Gauss-Markov theorem that is widely known in statistics.
Here we describe the ME-SQEC in the $q$ quadrature, where the CZ gate between the data qubit and ancilla qubit are performed, and then the ancilla qubit is measured.
Considering the Gauss-Markov theorem, the true deviation of the GKP qubit obeys the posterior probability corresponding to Gaussian distribution of mean $\delta$ and the variance ${\sigma '}^{2}_{q,{\rm D}}$, where $\delta$ and ${\sigma '}^{2}_{q,{\rm D}}$ are given by
\begin{equation}
\delta=\frac{\sigma^{2}_{q,{\rm D}}}{\sigma^{2}_{q,{\rm D}}+\sigma^{2}_{p,{\rm A}}}(\overline{\Delta}_{\rm p,A}- \overline{\Delta}_{\rm q,D})
\end{equation}
and
\begin{equation}
{\sigma '}^{2}_{q,{\rm D}}= \frac{\sigma^{2}_{q,{\rm D}}\sigma^{2}_{p,{\rm A}}}{\sigma^{2}_{q,{\rm D}}+\sigma^{2}_{p,{\rm A}}}~,
\end{equation}
respectively.
We note that the proposed maximum-likelihood estimation is based on the fact that the true deviation values are obeyed the Gaussian distribution independently. 
Then, by performing the displacement operation on the data qubit by $\delta$, 
the variance of the qubit D in the $q$ quadrature reduces from $\sigma^{2}_{p,{\rm A}}$ to ${\sigma '}^{2}_{q,{\rm D}}$, while the variance in $p$ quadrature increases from $\sigma^{2}_{p,{\rm D}}$ to $\sigma^{2}_{p,{\rm D}}+\sigma^{2}_{q,{\rm A}}$.
In the case where $\sigma^{2}_{p,{\rm D}}=\sigma^{2}_{q,{\rm A}}=\sigma^{2}$, the ME-SQEC improves the variance of the data qubit in the $q$ quadrature by $\sigma^{2}/2$ in comparison to the SQEC without a maximum-likelihood estimation.
In the measurement, the qubit level error occurs when the deviation value is more than $\sqrt{\pi}/2$ and the misidentification of the bit value occurs.
To reduce the probability of misidentifying the bit value of the ancilla, we use the HRM during the construction of the small-scale cluster state, as described in the next section.

 \section{Highly-reliable large-scale cluster state construction}\label{Sec3}

In this section, we introduce the scheme to apply the ME-SQEC to the construction of a small-scale cluster state.
Then we describe the large-scale cluster state construction from the small-scale cluster states, where the repetition code with the analog QEC is employed to reduce the measurement errors.

\subsection{Small-scale cluster state construction}\label{Sec3A}
 
\begin{figure}[t]
\begin{center}
 \includegraphics[scale=1.0]{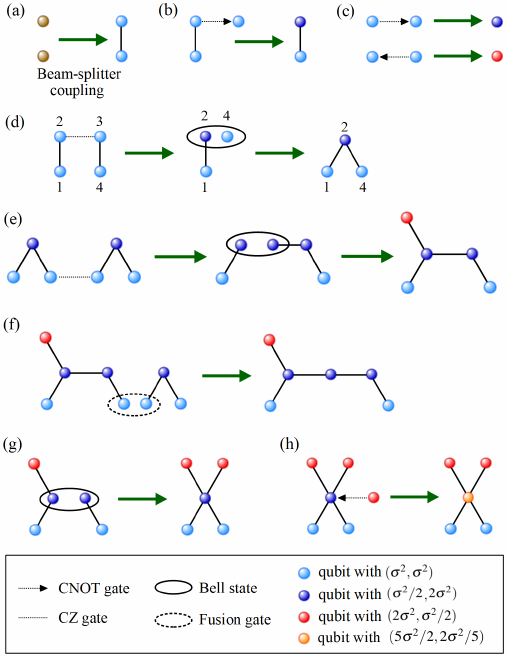} 
\setlength\abovecaptionskip{0pt}
\caption{The preparation of the small-scale cluster states with the ME-SQEC and the highly-reliable measurement (HRM).
(a) The generation of the balanced-entangled pair of two GKP qubits via 50:50 beam-splitter coupling between two qunaught states.
(b) The generation of the biased-single qubits via the ME-SQEC.
(c) The generation of the biased-entangled pair via the ME-SQEC in the $p$ quadrature on one of the biased-entangled pair in Fig.~\ref{fig2}(b), where the ancilla qubit is prepared in $\ket {\widetilde{0}}$ as the controlled qubit for the CNOT gate.
(d) The generation of the 3-tree cluster state from the two balanced-entangled pairs.
(e) The construction of the 5-mode cluster state from the two 3-tree cluster states.
(f) The construction of the 6-mode cluster state from the 5-mode cluster state and the 3-tree cluster state.
(g)(h) The construction of two types of 5-tree cluster states.
}
\label{fig2}
\end{center}
\end{figure}

In this work, we prepare the small-scale cluster states from the qunaught states and the GKP qubit with HRM and ME-SQEC.
Figure~\ref{fig2} shows the schematic diagram for the proposed scheme to prepare the reliable small-scale cluster state, where larger scale cluster states are generated from smaller scale states via the fusion gate and the two-mode gate with the HRM.
In the following, we see the construction of two types of the 5-tree cluster states described in Fig.~\ref{fig2}(g) and (h), which are used to construct larger cluster states with the HRM.

Firstly, we prepare two types of the two-mode entangled pairs, e.g., the balanced-entangled pair and the biased-entangled pair.
For the balanced-entangled pair, the Bell pair of GKP qubits are generated using a 50:50 beam splitter between the two qunaught states, as shown in Fig.~\ref{fig2}(a), where the qunaught state, $\ket{\varnothing}$, is described in Sec.~\ref{Sec2A}. 
For simplicity, we assume that the two-mode gate for the ME-SQEC is assumed to be ideal in this section.
For the variances, each GKP qubit of the entangled state has the variances $(\sigma^2, \sigma^2)$, i.e., variances in the $q$ and $p$ quadratures are both $\sigma^2$. 
After the Fourier transformation on one of the qubits, we obtain the two-mode cluster state.
For the biased-entangled pairs, the ME-SQEC with the HRM in the $q$ quadrature is applied to one of the entangled pair, as shown in Fig.~\ref{fig2}(b), where the variances are transformed as $(\sigma^2, \sigma^2)\mapsto (\sigma^2/2, 2\sigma^2)$. 
Figure~\ref{fig2}(c) shows the preparation of the biased-single qubit via the ME-SQECs in the $q$ and $p$ quadratures, where the variances are transformed from $(\sigma^2, \sigma^2)$ to $(\sigma^2/2, 2\sigma^2)$ and $(2\sigma^2, \sigma^2/2)$, respectively. The two biased-single qubits are used for the ME-SQECs on the qubit whose variance is not $(\sigma^2, \sigma^2)$.

Then, we generate the 3-tree cluster states via the CZ gate between each qubit of the two balanced-entangled pairs, as shown in Fig.~\ref{fig2}(d)
Considering the variances of the GKP qubit in the balanced-entangled pair, the CZ gate transforms the variances of the qubits 2 and 3 as  $(\sigma^2, \sigma^2)\mapsto (\sigma^2, 2\sigma^2)$.
After the measurement of the qubit 3, the qubits 2 and 4 are entangled as the Bell state. After Fourier transformation on the qubit 4, the 3-tree cluster state is prepared, where the variances for the qubit 1(4) and for the qubit 3 are $(\sigma^2, \sigma^2)$ and $(\sigma^2/2, 2\sigma^2)$, respectively. We note that the ME-SQEC is employed the measurement of the qubit 3 to reduce the variance in the $q$ quadrature.
Figure~\ref{fig2}(e) shows the construction of the 5-mode cluster state, where the two 3-tree cluster states are entangled by the CZ gate and one of the qubits is measured. In the measurement, the ME-SQEC with the HRM is employed on the qubit whose variances are transformed as $(\sigma^2, \sigma^2)\mapsto (\sigma^2/2, 2\sigma^2)$. After the Fourier transformation, the 5-mode cluster state is prepared, where the Fourier transformation transforms variances as $(\sigma^2/2, 2\sigma^2)\mapsto (2\sigma^2, 2\sigma^2/2)$.
Figure~\ref{fig2}(f) shows the construction of the 6-mode cluster state from the 5-mode cluster state and the 3-tree cluster state via the Bell measurement.
We call the entanglement generation via the Bell measurement as the fusion gate. 

Then, we prepare two types of 5-tree cluster states, as shown in Fig.~\ref{fig2}(g) and (h).
The 5-tree cluster state in Fig.~\ref{fig2}(g) is prepared by the measurement of one of the qubits in the 6-mode cluster state and the Fourier transformation.
We note that the error probability in the measurement during the construction of the cluster states in Fig.\ref{fig2}(d)-(g) is given by $E_{v_{\rm up}}(2\sigma^2) $.
The 5-tree cluster state in Fig.~\ref{fig2}(h) is prepared by the ME-SQEC in the $p$ quadrature between the two qubits in the 5-tree cluster state in Fig.~\ref{fig2}(g) and the biased-single qubit in Fig.~\ref{fig2}(c). This ME-SQEC transforms the variance as $(\sigma^2/2, 2\sigma^2)\mapsto (5\sigma^2/2, 2\sigma^2/5)$.
The error probability in this ME-SQEC is given by $E_{v_{\rm up}}(5\sigma^2/2)$.

\begin{figure*}[t]
 \includegraphics[angle=0, width=2.05\columnwidth]{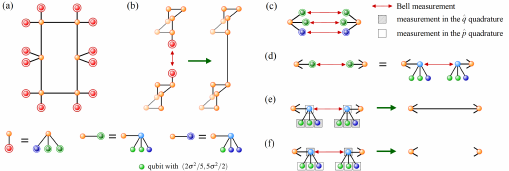} 
\setlength\abovecaptionskip{10pt}
  \caption{The construction of a large 3D cluster state. 
(a) The hexagonal cluster state with the encoded leaf qubits, where each of six node qubits is entangled with 2$L$ encoded leaf qubits and two neighboring node qubits.
Each encoded leaf qubit consists of a single leaf qubit and $m$ ancilla qubits. 
(b) The encoded Bell measurement between the two encoded leaf qubits in the neighboring hexagonal cluster states.
(c) The $m$-Bell measurements for each encoded Bell measurement.
(d) Each measurement of the $m$ Bell measurements between the two neighboring qubits.
(e) The measurement on the ancillae in the $q$ quadrature to keep the most reliable entanglement by comparing the reliabilities obtained from the measurement results of the Bell measurement. 
(f) The measurement on the ancillae in the $p$ quadrature to discard the entanglement except for the most reliable entanglement, which corresponds to the encoded measurement with the $m$-repetition code in the $q$ quadrature.
}
\label{fig3}
\end{figure*}
After the preparation of the small-scale cluster states described in Fig.~\ref{fig2}, we construct the encoded hexagonal cluster state.
Figure~\ref{fig3}(a) shows the encoded hexagonal cluster state, where each qubit of six node qubits is connected with the two encoded leaf qubits.
Each encoded leaf qubit consists of the encoded ancilla qubits, and each encoded ancilla qubit consists of the single leaf qubit and three ancilla qubits. 
The procedure for generating such small-scale cluster states is similar to that introduced in Ref.~\cite{fukui2018high}.
The detail for the construction of the encoded hexagonal cluster is described in Appendix~\ref{AppB}.
In the construction of the hexagonal cluster state, the maximum error probability for the fusion gate with the proposed scheme is $E_{v_{\rm up}}(5\sigma^2/2) $, while that with the conventional scheme~\cite{fukui2018high} is $E_{v_{\rm up}}(3\sigma^2)$.
In addition, the error probabilities for the ME-SQECs during the small-scale cluster states are limited to $E_{v_{\rm up}}(5\sigma^2/2) $ at most.
Accordingly, we obtain the 3-tree cluster state with a low error accumulation.

\subsection{Large-scale cluster state construction with the encoded measurement}\label{Sec3B}
After the preparation of the encoded hexagonal cluster states, we construct the so-called Raussendorf--Harrington--Goyal lattice referred to as the 3D cluster state in this work.
In this step, the large-scale 3D cluster state must be constructed deterministically from the encoded hexagonal cluster states since large-scale QC should be implemented deterministically.
In Ref.~\cite{fukui2018high}, the large-scale 3D cluster state is constructed from the hexagonal cluster described by using the deterministic fusion gate without the HRM.
The bottleneck to improve the threshold of the squeezing level in Ref.~\cite{fukui2018high} is the error derived from the deterministic fusion gate since the error probability of the deterministic fusion gate is several orders of magnitude higher than the fusion gate with the HRM.
In this work, we employ the encoded hexagonal cluster state to reduce the error probability of the deterministic fusion gate by using encoded ancilla qubits, as shown in Fig.~\ref{fig3}(b).
Consequently, we can perform topologically protected MBQC on the highly-reliable 3D cluster state.

Here we describe the fusion gate with the encoded measurement in more detail, which realizes the deterministic and reliable entanglement generation between neighboring node qubits.
Figures~\ref{fig3}(c)--(f) shows the schematic diagram for the encoded measurement, where there are three steps.
Firstly, we implement the Bell measurement between the two leaf qubits of neighboring hexagonal clusters, as shown in Fig.~\ref{fig3}(d).
Then, we select the most reliable entanglement from the measurement results by comparing $L$ likelihoods for the Bell measurements.
The likelihoods are obtained from measurement results as follows.
We assume that the $i$-the measurement deviations $\Delta_{{\rm m, A}_i}$ and $\Delta_{{\rm m, B}_i}$ are obtained from the $i$-the Bell measurement on the two $i$-the leaf qubits whose variance in the $q(p)$ quadrature is $\sigma^2_{{\rm leaf},q(p)}$.
The $i$-th likelihood is calculated by 
\begin{equation}
F_{i}=f(\Delta_{{\rm m, A}_i})f(\Delta_{{\rm m,B}_i}), 
\end{equation}
where $f(x)$ obeys the Gaussian distribution with mean zero and variance $\sigma^2_{{\rm leaf},q}+\sigma^2_{{\rm leaf},p}=2\sigma^2$, i.e. the sum of the variances for the leaf qubit in the $q$ and $p$ quadratures. 

Thirdly, we keep the most reliable entanglement, while we remove the entanglement except for the most reliable one by comparing the $m$ likelihoods.
To keep the entanglement, we measure the ancillae connected with the most reliable entanglement in the $q$ quadrature, as shown in Fig.~\ref{fig3}(e).
The probability of misidentifying the bit value in the $q$ quadrature is sufficiently small compared with that in the $p$ quadrature, e.g. the variances in the $q$ are $\sigma^2/2$ or $2\sigma^2/5$, while the variance in the $p$ quadrature are $2\sigma^2$ or $5\sigma^2/2$.
As a result, the ancillae are removed from the entanglement with a low error probability.
For the entanglement except for the most reliable entanglement, we measure encoded measurement in the $q$ quadrature. In the encoded measurement, we measure the ancillae in the $p$ quadrature so that we obtain the bit value of the leaf qubit in the $q$ quadrature, as shown in Fig.~\ref{fig3}(f), where the measurements of ancillae is encoded by the $m$-repetition code.
In the $m$-repetition code for the encoded leaf qubit without error, the node qubit labeled by ${\rm N}$ and the $i$-th ancilla labeled by ${\rm A}_{i}$ are stabilized by 
\begin{equation}
\hat{Z}_{{\rm N}}\hat{X}_{{\rm A}_i}=+1~(i=1,2,\cdots, m), 
\end{equation}
where $m$ corresponds to the number of ancilla qubits. After the measurements of the $m$ ancilla qubits, we implement a majority vote among the measurement results of ancillae in the $p$ quadrature.
As an example, we consider the case that the measurement results for the ancillae with $m=3$ are $\hat{X}_{{\rm A}_1}=+1$ , $\hat{X}_{{\rm A}_2 }=+1$,  and $\hat{X}_{{\rm A}_3}=-1$, i.e., the bit values of the qubits ${\rm A}_1$, ${\rm A}_2$, and ${\rm A}_3$ are measured as 0, 0, and 1, respectively.
In this case, the bit value for the node qubit in the $q$ quadrature is determined to be 0 by a majority vote.
This type of the indirect measurement via ancilla qubits was introduced by Varnava $et$ $al.$~\cite{varnava2006loss} for loss-tolerant optical quantum information processing.
In our method, the analog QEC~\cite{fukui2017analog} is applied to the repetition code to enhance the QEC performance, where we compare two likelihoods for the logical bit values from the measurement results of the $m$ ancilla qubits. The procedure for the $m$-repetition code with the analog QEC is described in Ref.~\cite{fukui2017analog}.
Consequently, we can obtain the large-scale 3D cluster state with a low error accumulation via the deterministic fusion gate with the encoded measurement.

\begin{figure}[t]
\begin{center}
 \includegraphics[angle=0, width=1.0\columnwidth]{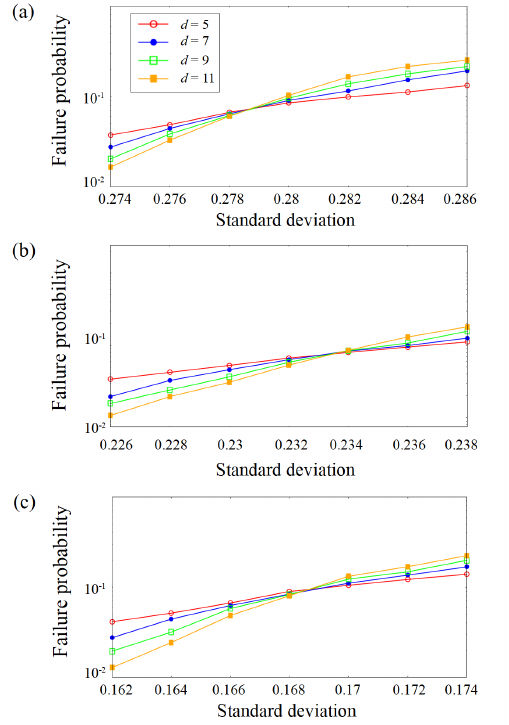} 
\setlength\abovecaptionskip{0pt}
\caption{Simulation results of the topologically protected MBQC with the analog QEC for the code distances $d$ = 5, 7, 9, and 11.
The failure probabilities of the QEC are plotted as a function of the standard deviation of the GKP qubits for loss in the homodyne measurement, (a)~$0$~$\%$, (b) $5$~$\% $, and (c) $10$~$\%$, respectively.
The QEC process is simulated by using the 3D cluster prepared by the proposed method with $v_{\rm up}=9\sqrt{\pi}/20$.
The simulation results are obtained from 10000 samples.}
\label{fig4}
\end{center}
\end{figure}
 \section{Threshold calculation}\label{Sec4}
In this section, we numerically calculate the threshold of the squeezing level required for FTQC.
In the numerical calculation, we simulate the QEC process for topologically protected MBQC by using the minimum-weight perfect-matching algorithm~\cite{edmonds1965paths,kolmogorov2009blossom} for the code distances $d$ = 5, 7, 9, and 11. 
In the simulation, we set the upper limit $v_{\rm up}$ for the HRM to $9\sqrt{\pi}/20$, and the number of ancilla qubits for the encoded measurement in the deterministic fusion gate is set to $m=3$.
In addition to the repetition code in the deterministic fusion gate, we apply the analog QEC~\cite{fukui2017analog} to decoding for the surface code in topologically protected MBQC to improve the QEC performance. The procedure to apply the analog QEC to the surface code has been described in Ref.~\cite{fukui2018high}. For an error model, we consider the errors derived from the variance for the node qubit itself, the accumulation of errors during the hexagonal cluster construction, and deterministic fusion gate, assuming noise in the two-mode gate and homodyne measurement. 
For the imperfect two-mode gate described in Sec.~\ref{Sec2B}, we assume that the squeezing level of the squeezed vacuum state is 15.0.

In Fig.~\ref{fig4}, the logical error probabilities are plotted as a function of the standard deviation for photon loss in the homodyne measurement $l = 0$, $5$, and $10$~$\%$, where photon loss $l$ is equal to $1-\eta$ with a transmittance coefficient $\eta$ described in Sec.~\ref{Sec2C}.
The numerical results confirm that our method for $l = 0$, $5$, and $10$~$\%$ achieves the threshold values of the standard deviation, around 0.278, 0.234, and 0.168, which correspond to the threshold values of the squeezing level, around 8.1, 9.6, and 12.4 dB, respectively.
The proposed scheme improves the threshold by $\sim1.8$ dB compared to the previous one in Ref.~\cite{fukui2018high}
Thus, our scheme provides high-threshold FTQC under noise in homodyene measurement and the two-mode gate.

\begin{figure*}[t]
 \includegraphics[angle=0, width=2.0\columnwidth]{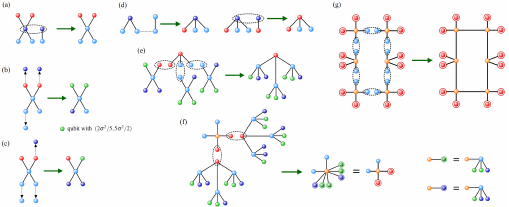} 
    \setlength\abovecaptionskip{10pt}
\caption{The construction of the encoded hexagonal cluster state via fusion gates with the HRM.
(a)(b)(c) The construction of three types of the 5-tree cluster state.
(d) The construction of the 4-tree cluster state.
(e) The construction of the encoded leaf qubit from the 4-tree cluster state and the 5-tree cluster states.
(f) The construction of the encoded 5-tree cluster state consisting of three qubits and two encoded qubits. Each encoded qubit consists of the $L=3$ encoded leaf qubits. Each encoded leaf qubits consists of the $m+1=4$ qubits used for the $m$-repetition code.
(g) The construction of the encoded hexagonal cluster with the encoded leaf qubits from the six encoded 5-tree cluster states.
}
\label{fig1a}
\end{figure*}

\section{Discussion and conclusion}\label{Sec5}
In this work, we have developed the method to perform high-threshold FTQC with the GKP qubit under noise in the two-mode gate and the homodyne measurement.
In our method, we have proposed a maximum-likelihood method to reduce a noise of the GKP qubits in the SQEC and the deterministic fusion gate with the encoded measurement, and combined the proposed methods with the conventional high-threshold FTQC~\cite{fukui2018high}.
The numerical calculations have shown that the required squeezing level can be improved to less than 10 dB with analog QEC up to about the transmission loss in the homodyne measurement 5~$\%$, which could be able to generate with the near-term experimental setup
Furthermore, our method enables us to CV-FTQC with 12.4 dB for the transmission loss 10~$\%$, showing that CV-FTQC with the GKP qubits is robust against photon loss.
In addition, there has been experimental progress in the improvement of a squeezing level toward CV quantum information processing, e.g. squeezed light with an optical cavity~\cite{suzuki20067,nagashima2008generation, vahlbruch2016detection} and waveguide devices~\cite{kashiwazaki2020continuous,kashiwazaki2021fabrication, kashiwazaki2023over}.
Hence, we believe this work will open up a way toward CV-FTQC with a moderate squeezing level.

In future works, we will analyze the resource for CV-FTQC, e.g. the number of qubits required to realize a quantum algorithm at various squeezing levels.
In addition, we could apply the proposed scheme to noise including the imperfection during the generation of the GKP qubit, e.g., such a noise model has been studied in Refs.~\cite{douce2019probabilistic,shi2019fault}.

\section*{Acknowledgements}
KF thanks Keisuke Fujii for useful discussions. This work was partly supported by JST [Moonshot R$\&$D][Grant No. JPMJMS2064][Grant No. JPMJMS2061], UTokyo Foundation, and donations from Nichia Corporation. 
\appendix

\section{Single-qubit level QEC in the $p$ quadrature}\label{AppA}
We describe the SQEC in the $p$ quadrature after the SQEC in the $q$ quadrature.
The SQEC in the $p$ quadrature is performed using the additional ancilla qubit ${\rm A}_2$ prepared in the state $\ket {\widetilde{+}}_{{\rm A}_2}=(\ket {\widetilde{0}}_{{\rm A}_2}+\ket {\widetilde{1}}_{{\rm A}_2})/\sqrt{2}$.
The data qubit D is interacted with the ancilla ${\rm A}_2$ by the CX gate, where the ancilla qubit is assumed to be the control qubit.
Regarding the deviation, the CX gate operation transforms the deviations in the $q$ and $p$ quadratures as
\begin{eqnarray}
\overline{\Delta}_{\rm {\it q},a_2} &\to & \overline{\Delta}_{\rm {\it q},a_2},\\
 \overline{\Delta}_{\rm {\it p},a_2} &\to &\overline{\Delta}_{\rm {\it p},a_2}-\overline{\Delta}_{\rm {\it p},D}+\overline{\Delta}_{\rm {\it p},a}, \\
-\overline{\Delta}_{\rm {\it q},a}&\to  &-\overline{\Delta}_{\rm {\it q},a} + \overline{\Delta}_{\rm {\it q},a_2}, \\
\overline{\Delta}_{\rm {\it p},D}&-&\overline{\Delta}_{\rm {\it p},a}\to \overline{\Delta}_{\rm {\it p},D}-\overline{\Delta}_{\rm {\it p},a},
\end{eqnarray}
where $\overline{\Delta}_{\rm {\it q},a2} ( \overline{\Delta}_{\rm {\it p},a2})$ is the true deviation value of the ancilla ${\rm A}_2$ in the $q$ ($p$) quadrature.
Then we measure the ancilla in the $p$ quadrature, and obtain the deviation of the ancilla ${\Delta}_{\rm m{\it p}, a_2}$. 
Then we measure the ancilla qubit in the $p$ quadrature, and obtain the measurement outcome $m_{p,{\rm m}}=(2 t + k)\sqrt{\pi}+{\Delta}_{\rm m{\it p}, a_2}$ to minimize $|{\Delta}_{\rm m{\it p}, a_2}|$, where $k$ is the bit value and $t = 0, \pm 1, \pm 2,\cdots$.
If $|{\Delta}_{\rm  m{\it p}, a2}| = |\overline{\Delta}_{\rm {\it q},a2}- \overline{\Delta}_{\rm {\it q},D} + \overline{\Delta}_{\rm {\it p},a}|$ is less than $\sqrt{\pi}/2$, the true deviation value of the data qubit in the $q$ quadrature changes from $\overline{\Delta}_{\rm {\it q},D} -\overline{\Delta}_{\rm {\it q},a}$ to $\overline{\Delta}_{\rm {\it q},a2}$ after the displacement operation.
If $|{\Delta}_{\rm  m{\it q}, a2}|$ is more than $\sqrt{\pi}/2$, the phase-flip error occurs after the displacement operation. 
Consequently, the sequential SQECs in the $q$ and $p$ quadratures transform the variances of the data qubit in the $q$ and $p$ quadratures as ${{\sigma}_{{\rm D},q}^2}\to{{\sigma}_{{\rm a},q}^2}+{{\sigma}_{{\rm a_2},q}^2}$ and ${{\sigma}_{{\rm D},p}^2}\to{{\sigma}_{{\rm a_2},p}^2}$, respectively. 
In this work, we employ the ME-SQEC to reduce the variances, as described in Sec.~\ref{Sec2E}.

\section{Construction of the hexagonal cluster state}\label{AppB}
We describe the construction of the encoded hexagonal cluster state from the cluster states described in Fig.~\ref{fig2}, where larger scale cluster states are generated from smaller scale states by using the fusion gate with the HRM, as shown in Fig.~\ref{fig1a}.
We describe the preparation of the encoded hexagonal cluster state, where each node qubit consists of $2L=6$ encoded leaf qubits and each encoded leaf qubit consists of $m=3$ ancilla qubits.
We note that the fusion gate does not increase and the HRM to ensure reliability is used for the fusion gate until the encoded hexagonal cluster state is prepared.
During the construction of the encoded hexagonal cluster state, the error probabilities is limited to $E_{v_{\rm up}}(5\sigma^2/2) $ at most for each measurement.

Figure~\ref{fig1a}(a)--(c) shows the construction of three types of the 5-tree cluster states.
For the 5-tree cluster state in Fig.~\ref{fig1a}(a), the cluster state is prepared by the fusion gate between the qubits of cluster states in Fig.~\ref{fig2}(b) and (g), where the error probability in the fusion gate is given by $E_{v_{\rm up}}(5\sigma^2/2)$
Figure~\ref{fig1a}(b) describes the 5-tree cluster state via the three ME-SQECs in the $q$ quadrature on the qubits in Fig.~\ref{fig1a}(a), where the error probabilities in the ME-SQECs is given by $E_{v_{\rm up}}(2\sigma^2)$ and $E_{v_{\rm up}}(5\sigma^2/2)$.
The ME-SQECs in the $q$ quadrature transforms variances as $(2\sigma^2, \sigma^2/2)\mapsto (2\sigma^2/5, 5\sigma^2/2)$ and $(\sigma^2, \sigma^2)\mapsto (\sigma^2/2, 2\sigma^2)$. 
Figure~\ref{fig1a}(c) describes the 5-tree cluster state via the three ME-SQECs in the $q$ quadrature on the qubit in the 5-tree cluster state in Fig.~\ref{fig1a}(b).
Figure~\ref{fig1a}(d) describes the 4-tree cluster state. For this 4-tree cluster state, we firstly prepare the 4-tree cluster state with the node qubit whose variances is $ (\sigma^2/2, 2\sigma^2)$, where we perform the CZ gate between the qubits in the 3-tree cluster state and the balanced-entangled pair, and implement the Fourier transformation after the measurement. Then, we apply the fusion gate between the qubits in the 4-tree cluster state and the biased-entangled pair to obtain the 4-tree cluster state with the node qubit whose variances is $ (2\sigma^2, \sigma^2/2)$, where the error probability in the fusion gate is given by $E_{v_{\rm up}}(5\sigma^2/2)$.
Figure~\ref{fig1a}(e) describes the construction of the encoded leaf qubit from the 4-tree cluster state and the 5-tree cluster states by using the fusion gates.
Figure~\ref{fig1a}(f) describes the construction of the encoded 5-tree cluster state via the fusion gate between the qubits in the cluster states in Fig.~\ref{fig1a} (d) and Fig.~\ref{fig1a}(e).
The encoded 5-tree cluster state, i.e. the encoded leaf qubit, consists of the $2L$ encoded ancilla qubits, where each encoded ancilla qubit consists of the single leaf qubit and three ancilla qubits. These ancilla qubits are used for the encoded measurement to realize the reliable entanglement generation between the node qubits in the neighboring hexagonal cluster states, as describe in Sec.~\ref{Sec3B}.
Finally, the encoded hexagonal cluster state is constructed from the six encoded 5-tree cluster states, as shown in Fig.~\ref{fig1a}(g).
Regarding the error probabilities, the probability for the fusion gate between the qubits with variances $(\sigma^2, \sigma^2)$, the error probability is given by $E_{v_{\rm up}}(2\sigma^2)$, and that with variances $(\sigma^2/2, 2\sigma^2)$ is $E_{v_{\rm up}}(5\sigma^2/2)$.
Accordingly, the error probabilities of the fusion gate are limited to $E_{v_{\rm up}}(5\sigma^2/2) $ at most during the preparation of the encoded hexagonal cluster state.
In the construction of the hexagonal cluster state, the maximum error probability for the proposed scheme, $E_{v_{\rm up}}(5\sigma^2/2)$, is improved, compared with that for the scheme in Ref.~\cite{fukui2018high}, $E_{v_{\rm up}}(3\sigma^2)$.
Thus, we obtain the highly-reliable encoded hexagonal cluster states with a low error accumulation at the cost of the success probability of the HRM, which are used for the deterministic generation of large-scale cluster state required for FTQC.

\bibliography{ref.bib}

\end{document}